\documentclass{iau}
\usepackage{graphicx}

\title[Magnetic Fields in Low-mass Stars] 
{Magnetic Fields in Low-Mass Stars:\\ An Overview of Observational Biases}

\author[Ansgar Reiners]   
{Ansgar Reiners}

\affiliation{$^1$Institut f\"ur Astrophysik, Georg-August Universit\"at
  G\"ottingen, \\ 37077 G\"ottingen, Germany\\ email: {\tt Ansgar.Reiners@phys.uni-goettingen.de}}

\pubyear{2014}
\volume{xxx}  
\pagerange{xxx--xxx}
\jname{ Magnetic Fields Throughout Stellar Evolution}
\editors{A.C. Editor, B.D. Editor \& C.E. Editor, eds.}
\begin{document}

\maketitle

\begin{abstract}
  Stellar magnetic dynamos are driven by rotation, rapidly rotating stars
  produce stronger magnetic fields than slowly rotating stars do. The Zeeman
  effect is the most important indicator of magnetic fields, but Zeeman
  broadening must be disentangled from other broadening mechanisms, mainly
  rotation. The relations between rotation and magnetic field generation,
  between Doppler and Zeeman line broadening, and between rotation, stellar
  radius, and angular momentum evolution introduce several observational
  biases that affect our picture of stellar magnetism. In this overview, a few
  of these relations are explicitly shown, and the currently known
  distribution of field measurements is presented.
\end{abstract}

\firstsection 
\section{Introduction}

An important difference between massive and low-mass stars is the presence of
an outer convection zone. In this article, we distinguish between high- and
low-mass stars on the basis of the presence of an outer convective zone;
high-mass stars have no outer convective envelopes. While they may have
convective cores, radiative energy transport dominates in the outer zones of
these stars. Because dissipation timescales are long, strong magnetic fields
may survive there, but fields are not generated, and fields that may be
generated in the core find no easy way to the surface.

Low-mass stars, on the other hand, have outer convective envelopes in which
magnetic fields decay within only a few decades or centuries \cite[(Chabrier
\& K\"uker, 2006)]{ChabrierKueker06}, and where motion of ionized particles
apparently manage to generate strong magnetic fields as for example in the
Sun. The efficiency of magnetic field generation through a dynamo process
depends on several conditions, but the details of these are not well known
\cite[(e.g., Charbonneau, 2010)]{Charbonneau10}. The Sun is one anchor for our
models of stellar dynamos. It is probably a fairly common representative of
its type \cite[(Basri et al., 2013)]{Basri13}, but we know many stars that are
a lot younger and more active. These stars produce orders of magnitude more
non-thermal radiation (activity). The reason for this is probably their faster
rotation leading to enhanced dynamo action powering non-thermal heating of the
chromosphere and corona.

Observations of magnetic fields in stars other than the Sun require relatively
high data quality. More important, the signatures of magnetic fields must be
disentangled from other effects, which is often difficult because the
characteristic properties of low-mass stars evolve in time (and differently
for different stellar masses). In this article, I introduce the main
characters important for spectroscopic measurements of magnetic fields and
their interpretation, and I present the currently known distribution of field
measurement using different techniques. A more comprehensive review about
observations of low-mass star magnetic fields can be found in \cite[Reiners
(2012)]{LivRev12} where the data used here are also presented.
 
\section{Cast of Characters}

Four main characters are conspiring in our picture of stellar dynamos and
their spectroscopic observations. They are the following:

\subsection{Zeeman}

\begin{figure}
  \centering
  \mbox{
    \parbox[c]{7cm}{\vspace{0pt}\includegraphics[width=7cm]{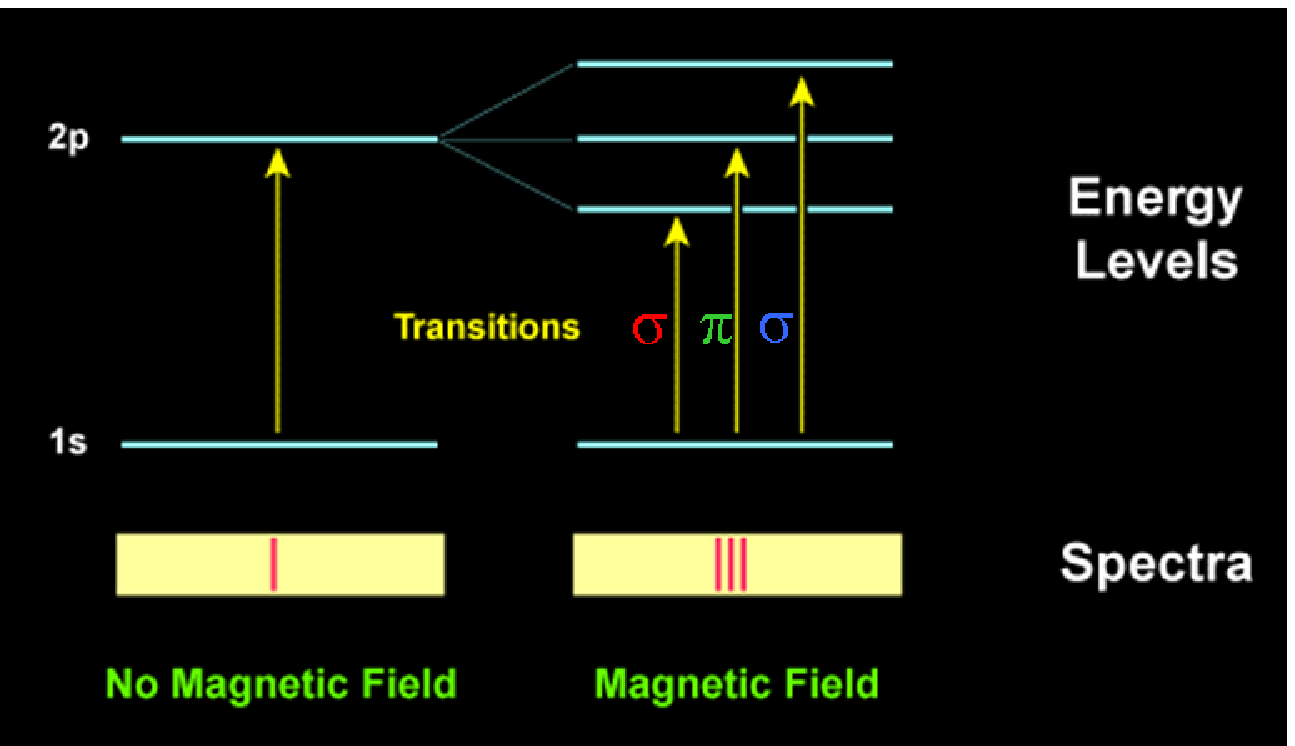}}\qquad
    \parbox[c]{5cm}{\vspace{0pt}\includegraphics[width=5cm]{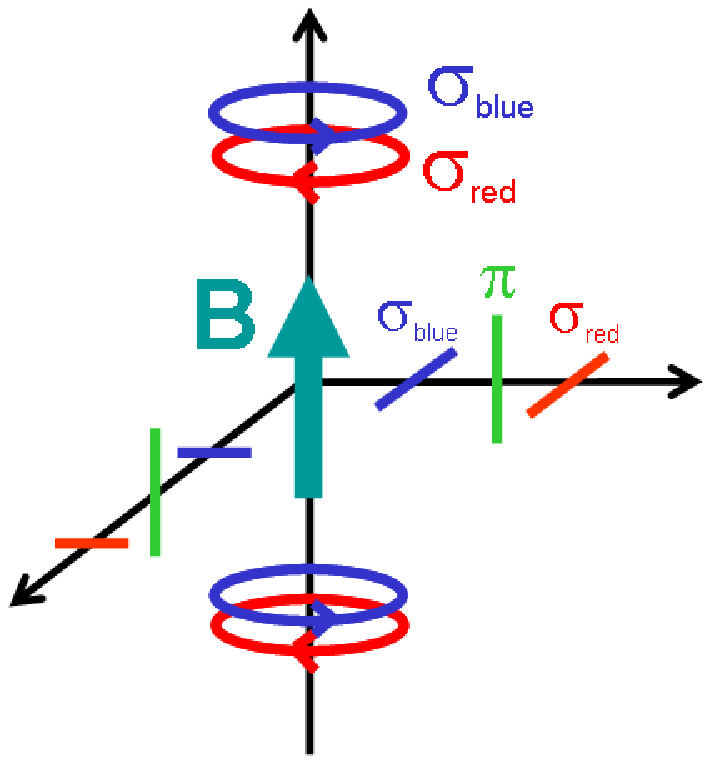}}}
  \caption{\label{fig:Zeeman}Simplified Zeeman splitting mechanism. The upper
    energy level is split into three levels in the presence of a magnetic
    field (\emph{right}). The three different components have different
    polarizations and produce very different signatures depending on the
    direction of observation (\emph{right}). }
\end{figure}  

The Zeeman effect is the most obvious and most direct consequence of the
presence of a magnetic field in a star. As we know from the Sun, magnetic
fields can lead to enhanced non-thermal radiation that we call activity, but
only the direct detection through the Zeeman effect can show that stars other
than the Sun really follow similar rules, and that other stars do indeed
produce average magnetic fields orders of magnitude stronger than the Sun
does.

The principle of the Zeeman effect is shown in Fig.\,\ref{fig:Zeeman}. The
energetic degeneracy between energy levels can be lifted by the presence of a
magnetic field, which typically leads to three different groups of
transitions, two $\sigma$-groups and one $\pi$-group. The groups have
different polarizations and are selectively emitted into certain directions
depending on the orientation of the magnetic field. The displacement of the
$\sigma$-groups with respect to the non-displaced $\pi$-group is

\begin{equation}
  \label{Eq:Zeemanl}
  \frac{\Delta \lambda}{\mathrm{m\AA}} = 46.67 \, g \left(\frac{\lambda_0}{\mathrm{\mu m}}\right)^2 \frac{B}{\mathrm{kG}}; 
\end{equation}
written in units of wavelength, the Zeeman effect is a function of
$\lambda^2$. In units of velocity, the Zeeman effect can be written as
\begin{equation}
  \label{Eq:Zeemanv}
  \frac{\Delta v}{\mathrm{m s^{-1}}} = 1.4 \, g \, \frac{\lambda_0}{\mathrm{\mu m}}
  \frac{B}{\mathrm G},
\end{equation}
which still depends on wavelength. At a wavelength of $\lambda = 1\,\mu$m, the
typical Zeeman displacement is $\Delta v = 1$\,m\,s$^{-1}$ for a field
strength of $B = 1$\,G.

\subsection{Stokes}

\begin{figure}
  \centering
      \includegraphics[width=.3\textwidth]{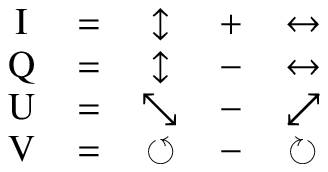}
  \caption{\label{fig:Stokes}The four Stokes parameters.}
\end{figure}

The polarization states of the $\pi$- and $\sigma$-components are
different. This provides great potential for the detection and measurement of
magnetic fields because different polarizations can be compared with each
other differentially. Individual polarization vectors, however, cannot simply
be observed but need to be filtered out, e.g., by the use of polarizing
beamsplitters \cite[(see, e.g., Tinbergen, 1996)]{Tinbergen96}. Together with
retarding waveplates, combinations of linear and circular polarizations can be
observed consecutively. One possible choice for observable combinations of
polarization states is defined by the so-called Stokes vectors, shown in
Fig.\,\ref{fig:Stokes}. Stokes~I is simply the integrated light, i.e. the sum
of of the two perpendicular linear or circular components. Stokes~Q and U are
the differences of the two perpendicular linear polarization components, the
reference frames of Q and U are rotated by $45^\circ$ with respect to each
other. Stokes~V is the difference between left- and right-handed circular
polarization.

Depending on the direction of observation, the linear and circular
polarization vectors carry different parts of the magnetic field
information. What is worse, regions of opposite polarity produce circular
polarization that can entirely cancel out each other. This is because the
blue-shifted circularly polarized component of a ``positive'' magnetic field
has exactly the same shift and amplitude as the analog component caused by a
``negative'' magnetic field, but the sign of that component is opposite. The
sum of the two Stokes~V components is therefore exactly zero.  Note that this
cancellation does not occur in the linearly polarized components because the
direction of polarization is identical for the two $\sigma$-components. It is
important to realize that the information in integrated light, Stokes~I,
depends on the direction of observation, too, mainly because the linearly
polarized $\pi$-components are invisible if the magnetic field direction is
parallel to the direction of observation.

\subsection{Doppler}

\begin{figure}
  \centering
      \includegraphics[width=\textwidth,bbllx=30,bblly=10,bburx=620,bbury=440]{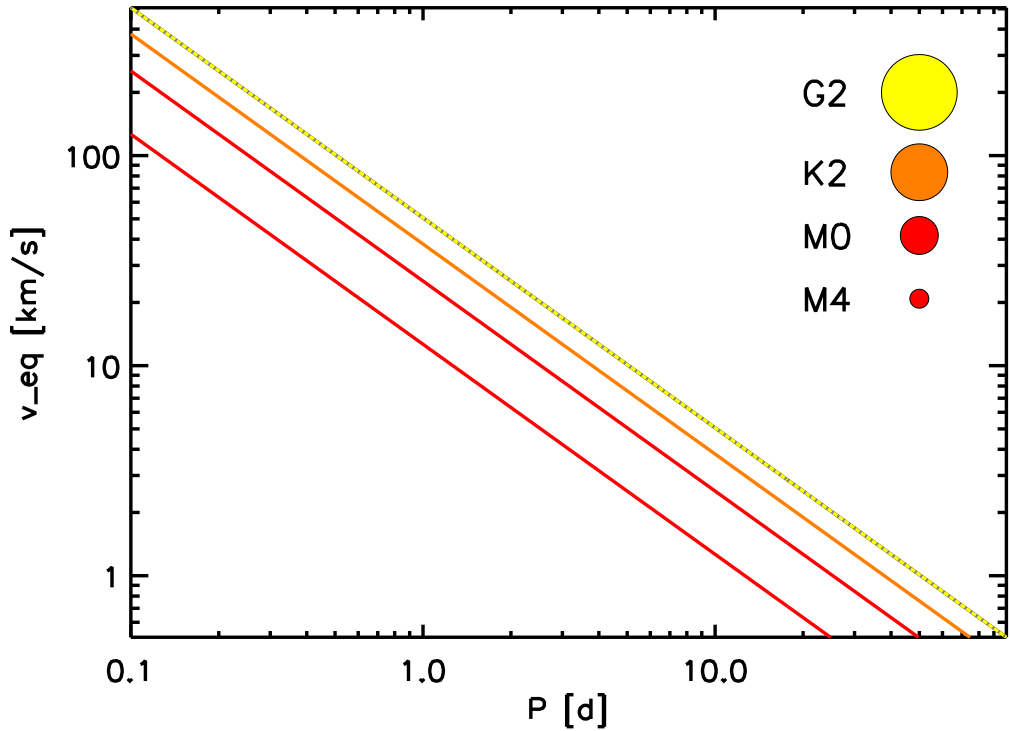}
      \caption{\label{fig:Doppler}Equatorial Doppler velocities for four MS
        stars as a function of their rotation periods. The spectral types and
        relative sizes are shown in the upper right legend. The vertical
        position of the relations follows the ordering in the legend (and
        colors of the lines match legend colors).}
\end{figure}

In a rotating star, light emitted from the side of the star that is
approaching the observer is blueshifted, and light from the other side of the
star is redshifted. This leads to net broadening of spectral lines and can be
used to determine the projected rotational velocity, $v\,\sin{i}$, of the star
\cite[(e.g., Gray, 2008)]{Gray08}. Low-mass stars as defined here (all stars
with outer convective envelopes) include all stars with masses and radii
between approximately 1.2 and 0.1 times the solar values, i.e., their
characteristic properties vary over more than one order of magnitude. Young
stars can also possess outer convective envelopes and are a lot larger than
main sequence (MS) stars adding to the great variety of targets. In
Fig.\,\ref{fig:Doppler}, the equatorial velocities of four different MS stars
are shown as a function of their rotation periods. For example, a G2 star with
a rotation period of $P = 10$\,d will have an equatorial velocity of
$v_{\mathrm{eq}} = 5$\,km\,s$^{-1}$, but an M4 star of the same period will
only show a maximum line broadening corresponding to the equatorial velocity
of $v_{\mathrm{eq}} = 1$\,km\,s$^{-1}$. This difference has severe
consequences for the observability of spectroscopic line diagnostics, as for
example Zeeman broadening, because Zeeman broadening must be disentangled from
Doppler broadening.

\subsection{Rossby}

\begin{figure}
  \centering
      \includegraphics[width=\textwidth,bbllx=30,bblly=10,bburx=620,bbury=440]{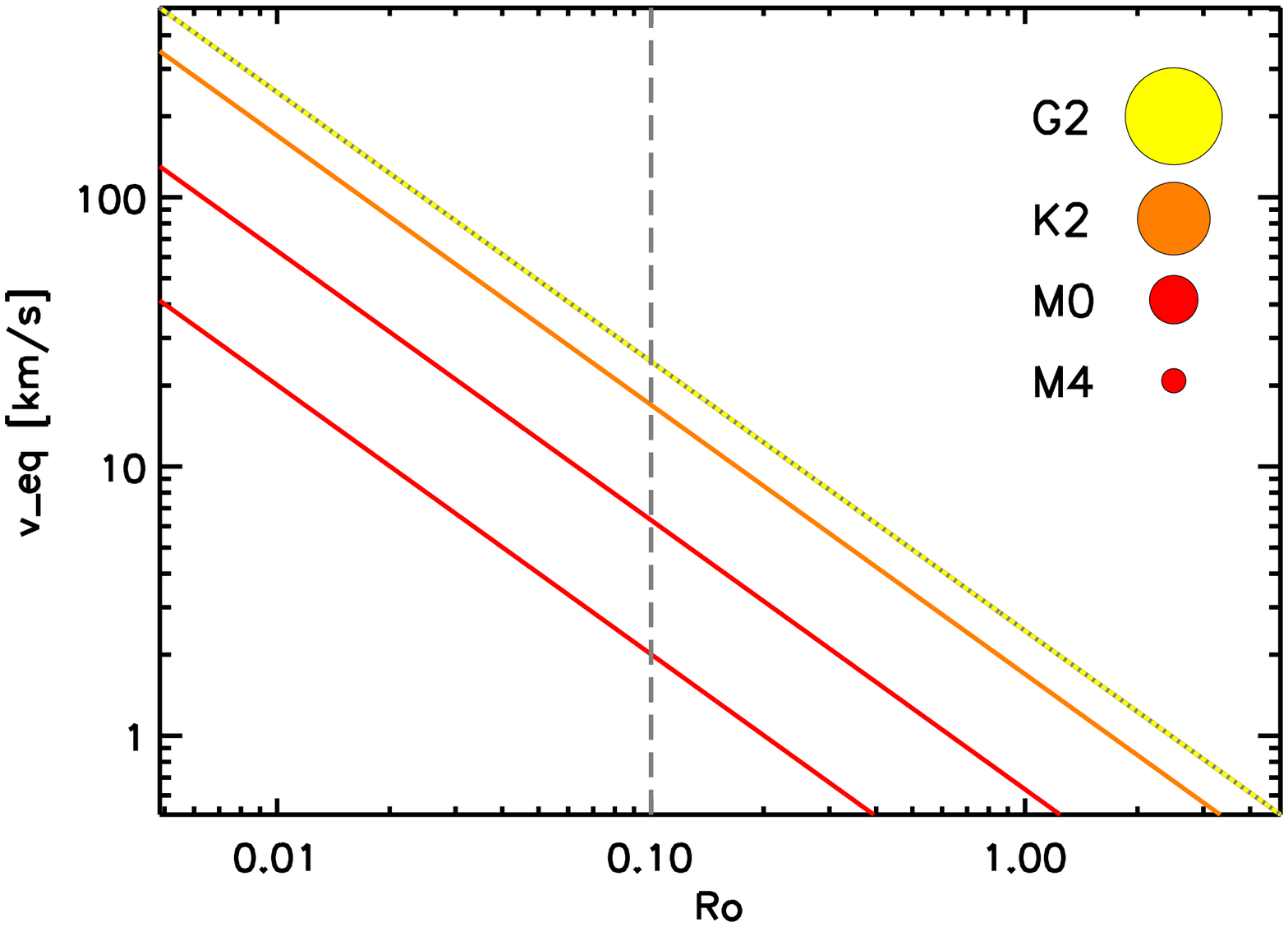}
      \caption{\label{fig:Rossby}Equatorial Doppler velocities for four MS
        stars as a function of Rossby number. The spectral types and relative
        sizes are shown in the upper right legend. The vertical position of
        the relations follows the ordering in the legend (and colors of the
        lines match legend colors).}
\end{figure}

A prediction from Dynamo theory is that the efficiency of a convective stellar
dynamo may depend on the ratio between Coriolis force and field dissipation
\cite[(e.g., Ossendrijver, 2003)]{Ossendrijver03}. This ratio can be expressed
in terms of typical convective and rotation timescales. For the rotation
timescale, an obvious choice is the rotational period. For the convective
timescale, a timescale used quite often is the convective overturn time, which
is defined as the typical convective velocity divided by the size of the
convection zone \cite[(Durney \& Latour, 1978)]{Durney78}. The ratio of the
two is called the Rossby number, $Ro = P/\tau_{\mathrm{conv}}$.

The convective overturn time is a slowly varying function of stellar mass, and
therefore the Rossby number is mostly determined by the value of the rotation
period. Nevertheless, if activity in different stars should be compared, it is
often useful to use the Rossby number instead of comparing rotation
periods. Figure\,\ref{fig:Rossby} shows the equatorial velocity on the surface
of a star as a function of Rossby number. It is similar as
Fig.\,\ref{fig:Doppler} but the differences in $v_{\mathrm{eq}}$ are even
larger for stars of different mass because not only the radius but also the
Rossby number is different.

\section{Zeeman or Doppler?}

\begin{figure}
  \centering
  \mbox{
    \parbox{.48\textwidth}{
      \includegraphics[width=.48\textwidth,bbllx=0,bblly=125,bburx=620,bbury=450]{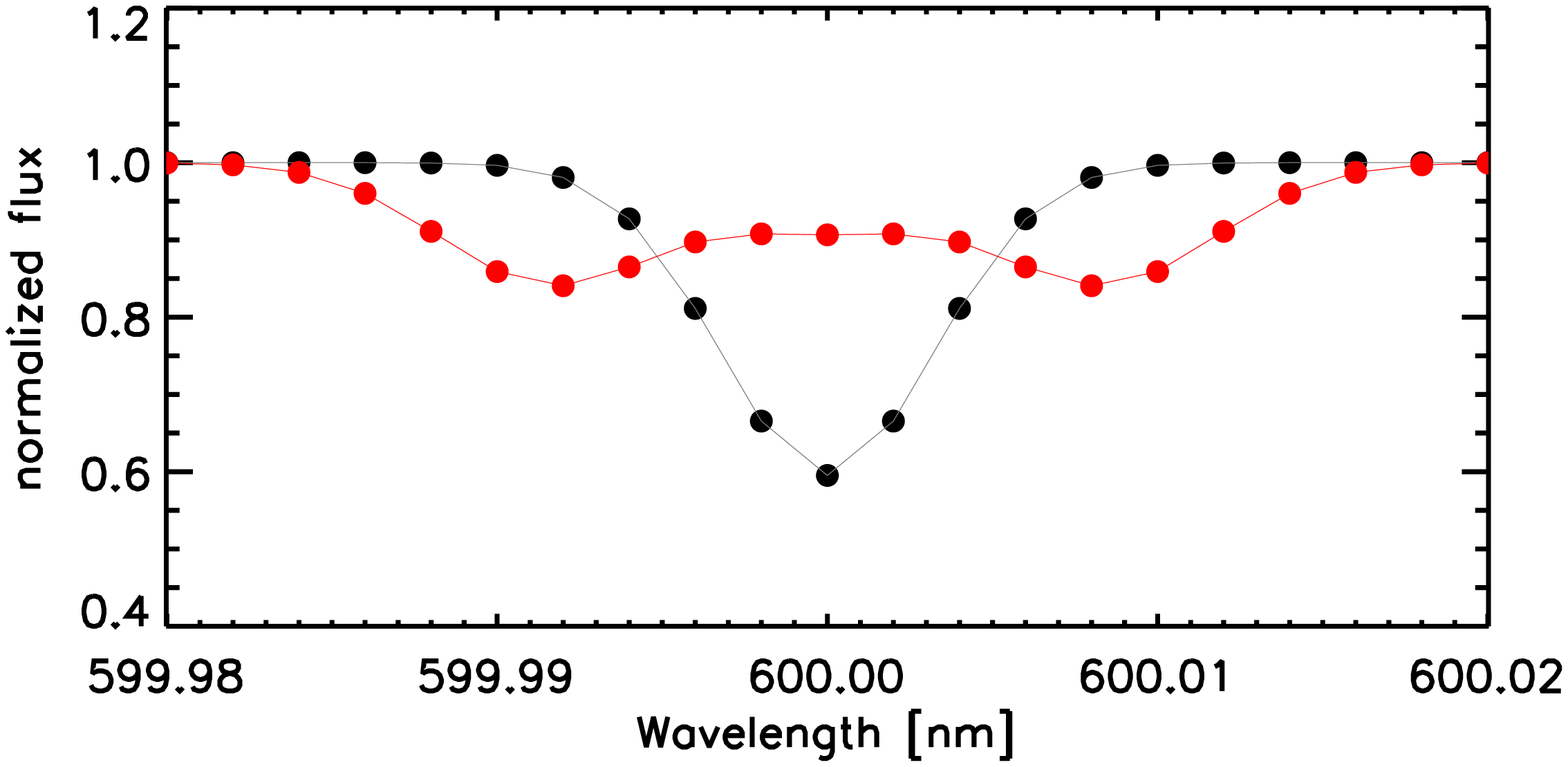}}
    \hspace{.04\textwidth}
    \parbox{.48\textwidth}{
      \includegraphics[width=.48\textwidth,bbllx=0,bblly=125,bburx=620,bbury=450]{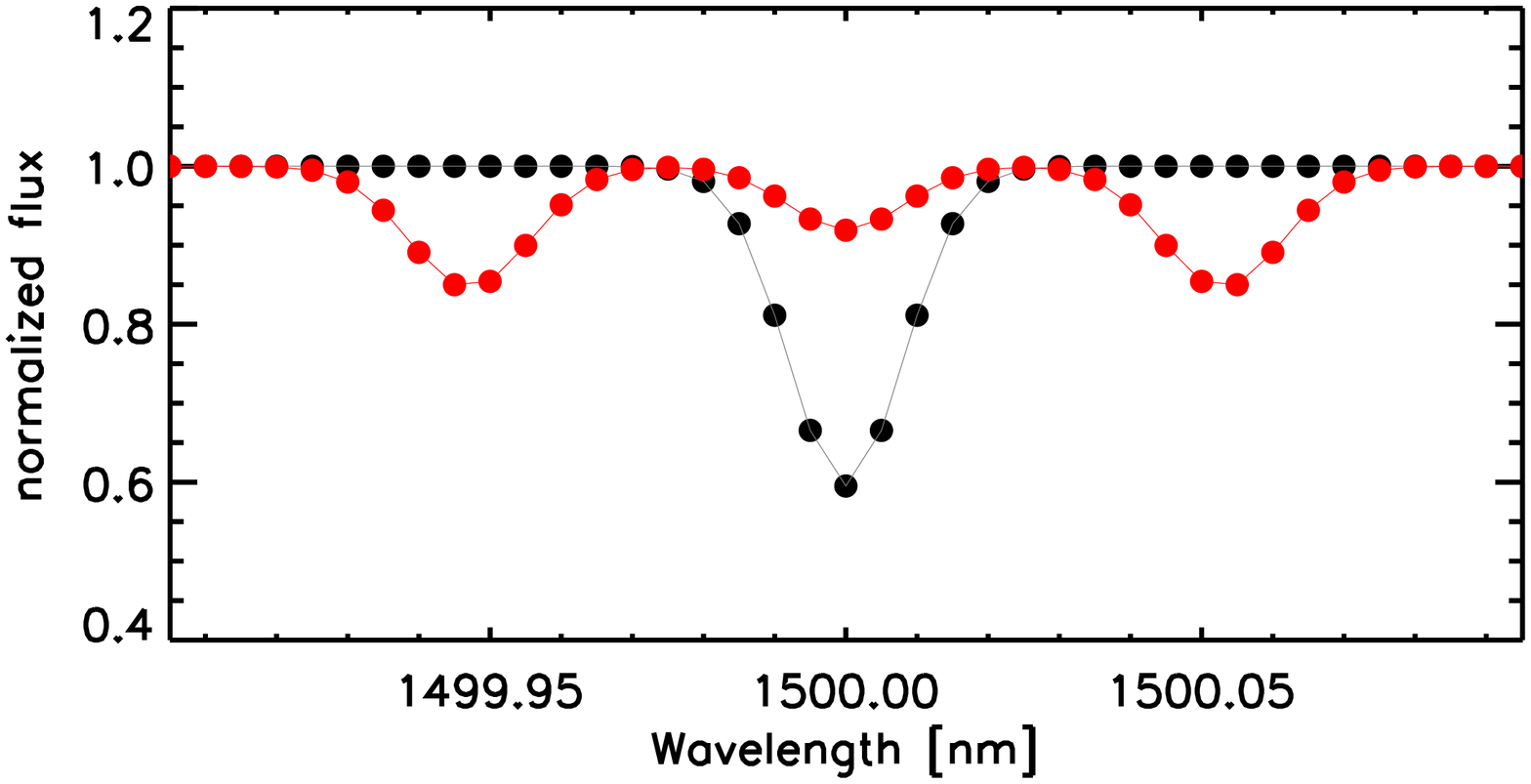}}
  }
  \mbox{
    \parbox{.48\textwidth}{
      \includegraphics[width=.48\textwidth,bbllx=0,bblly=125,bburx=620,bbury=450]{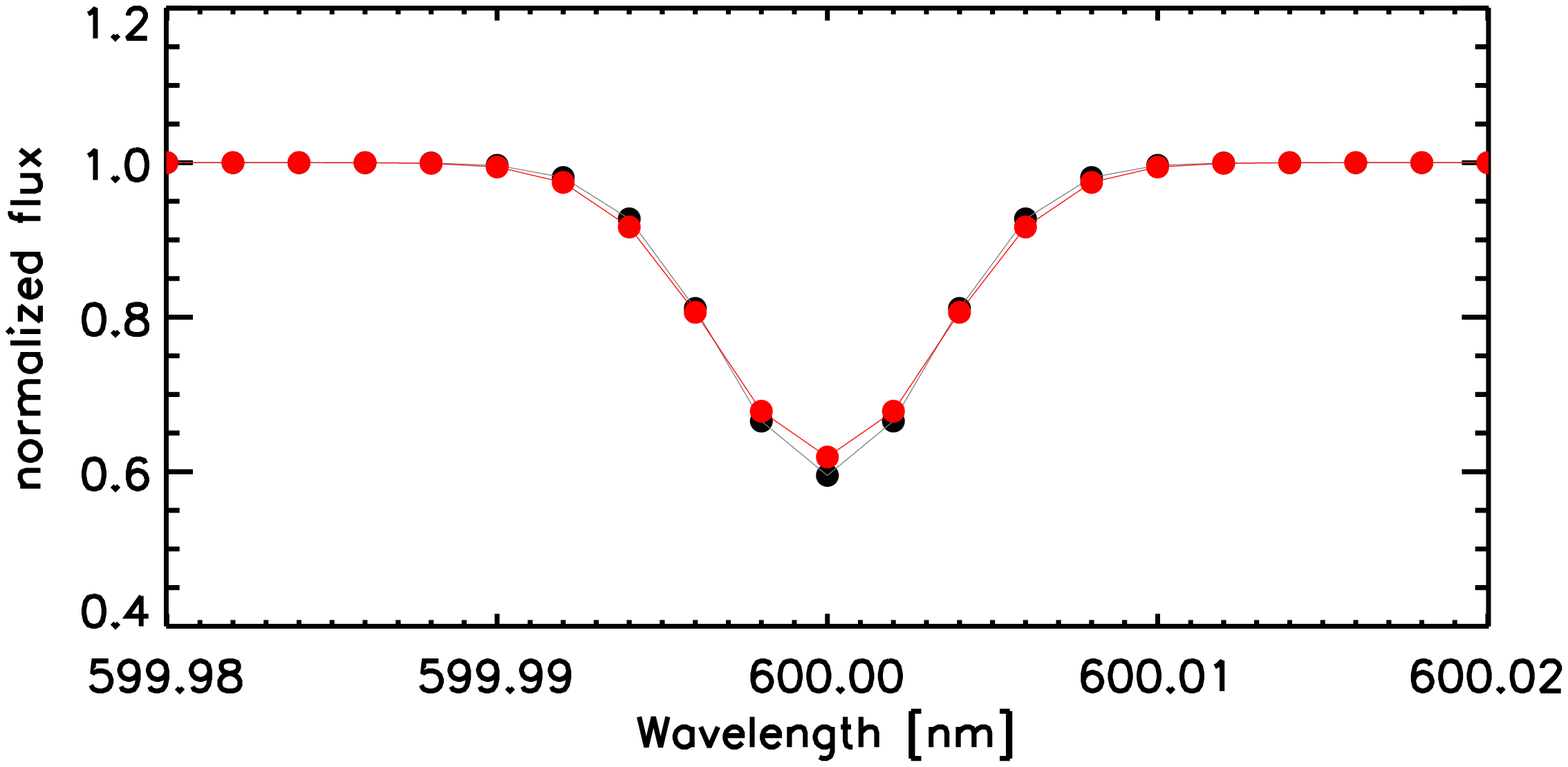}}
    \hspace{.04\textwidth}
    \parbox{.48\textwidth}{
      \includegraphics[width=.48\textwidth,bbllx=0,bblly=125,bburx=620,bbury=450]{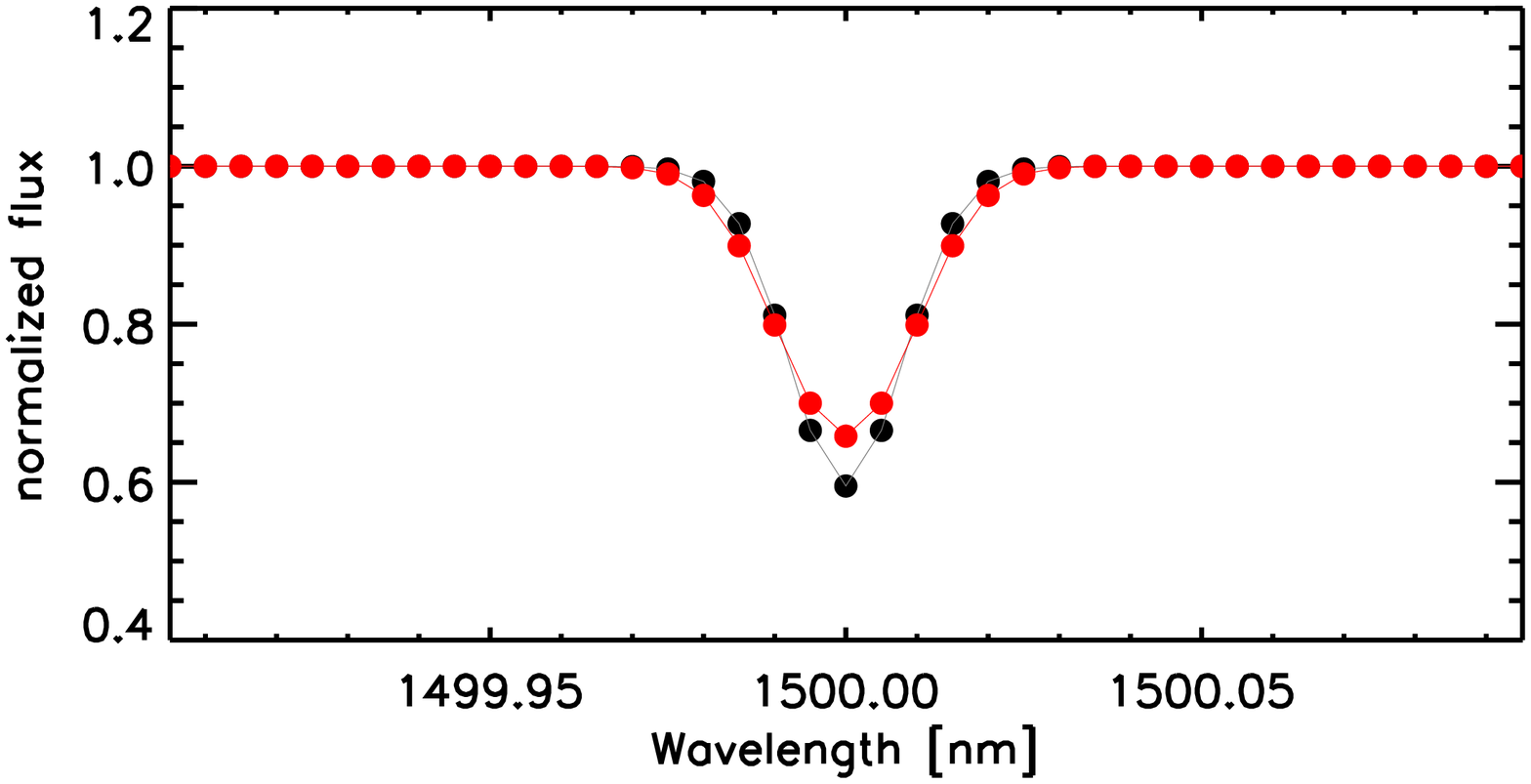}}
  }
  \mbox{
    \parbox{.48\textwidth}{
      \includegraphics[width=.48\textwidth,bbllx=0,bblly=125,bburx=620,bbury=450]{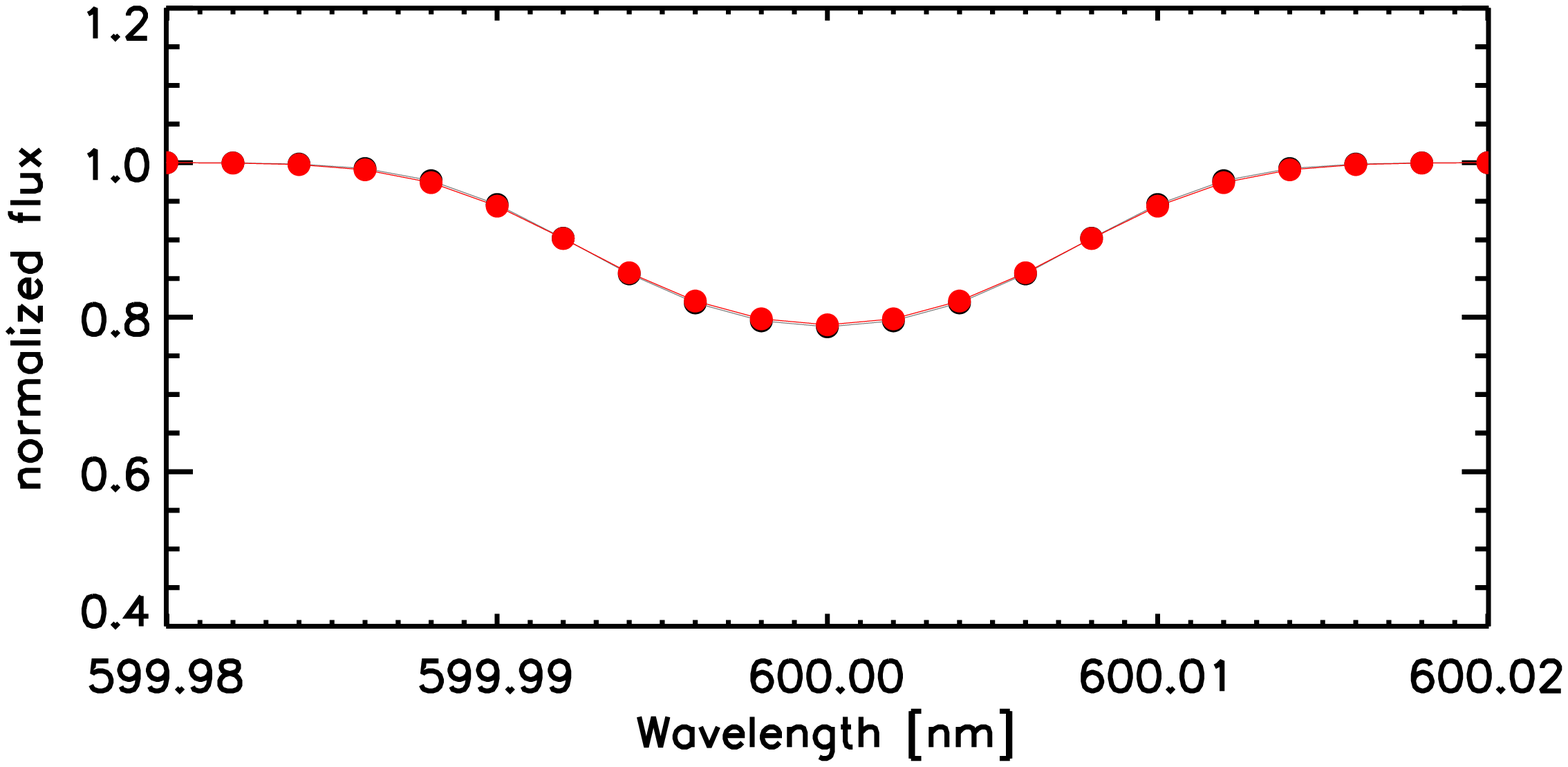}}
    \hspace{.04\textwidth}
    \parbox{.48\textwidth}{
      \includegraphics[width=.48\textwidth,bbllx=0,bblly=125,bburx=620,bbury=450]{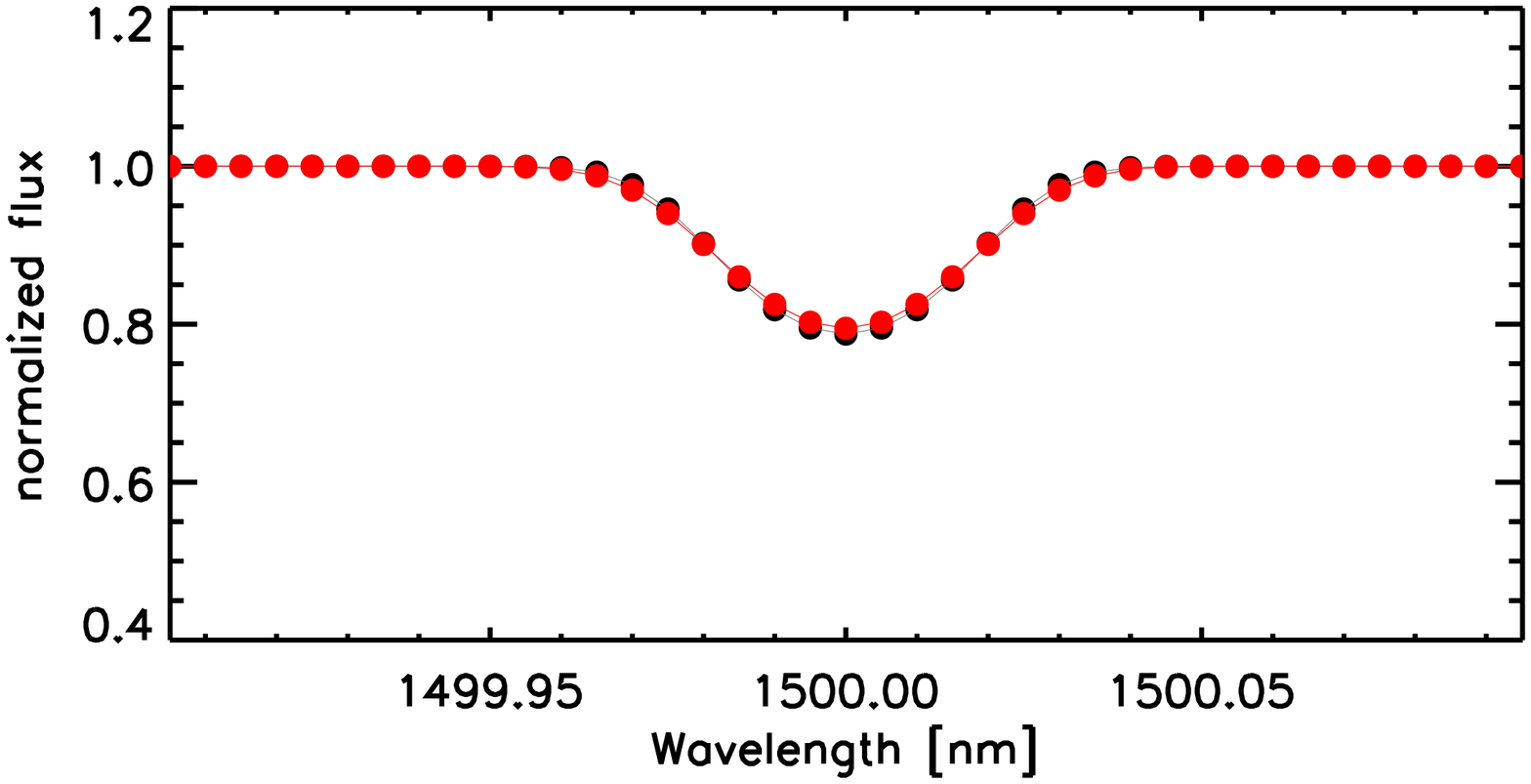}}
  }
  \caption{\label{fig:Spectra}The effect of Zeeman broadening on a single
    spectral line. \emph{Left:} A typical line at $\lambda = 600$\,nm;
    \emph{right:} a typical line at $\lambda = 1500$\,nm; $g = 2$ in all
    cases. \emph{Upper panel:} Zeeman broadening in a star with an average
    magnetic field of $B = 2000$\,G (red) compared to a line with no $B =
    0$\,G. \emph{Centre panel:} Effect of a field of $B = 200$\,G (red)
    compared to zero field strength. \emph{Bottom panel:} Effect of $B =
    200$\,G observed in a star rotating at $v\,\sin{i} = 5$\,km\,s$^{-1}$.}
\end{figure}

The relation between rotation and activity is a well-established observational
fact; slow rotators produce little activity, faster rotators produce more
\cite[(Pizzolato et al., 2003; Wright et al., 2011)]{Pizzolato03,
  Wright11}. The ratio between activity seen in non-thermal emission and the
star's bolometric luminosity is a function of the rotation period, but it
differs between different stars. For equal Rossby numbers, however, it is
expected that this ratio is similar for all stars. Therefore, convective
overturn times are sometimes motivated empirically by searching for the
function of $\tau_{\mathrm{conv}}$ that minimizes the scatter in the
activity-rotation relation \cite[(Noyes et al., 1984; Kiraga \& Stepien, 2007;
Wright et al., 2011)]{Noyes84, Kiraga07, Wright11}. A problem for the
theoretical calculations of $\tau_{\mathrm{conv}}$ is that it is not obvious
what definition of $\tau_{\mathrm{conv}}$ one should use -- is it the
convective overturn time at the bottom of the convective envelope, or the
weighted mean throughout the convection zone, or something different?

Measuring a magnetic field in a rotating star requires Zeeman
broadening to be a significant fraction of the total line broadening
that is dominated by rotation. In slow rotators ($Ro > 0.1$), the
magnetic field grows with rotation, in fast rotators, the field seems
to be saturated and the field does not grow further with rotation
\cite[(Reiners et al., 2009)]{Reiners09}. The ratio between Zeeman and
Doppler broadening is therefore smaller in the regime of saturated
dynamos.

The effect of Zeeman broadening in the presence of rotation is
displayed in Fig.\,\ref{fig:Spectra}; the bottom panel of this figure
shows a typical situation for a sun-like star at $Ro \approx 0.5$.

In the non-saturated part of the rotation-activity or rotation-magnetic field
relation, Zeeman broadening ($\Delta v_{\mathrm{Zeeman}}$) is approximately
proportional to Doppler broadening ($v_{\mathrm{eq}}$). Based on the
heterogeneous sample of Zeeman measurements collected in \cite[Reiners
(2012)]{LivRev12}, an estimate of the ratio for G dwarfs is
\begin{equation}
  \frac{\Delta v_{\mathrm{Zeeman}}}{v_{\mathrm{eq}}} \approx 0.07 \left( \frac{\lambda_{0}}{1\,\mathrm{\mu m}} \right) g.
\end{equation}

This ratio is approximately valid for all stars with non-saturated
activity. In more rapidly rotating stars, the ratio is smaller because
rotational broadening is larger but Zeeman broadening is saturated.

\section{Stokes' Choices}

\begin{figure}
  \centering
  \mbox{
    \parbox{.48\textwidth}{\hspace{.8cm} \textbf{G and K stars}}
    \hspace{.04\textwidth}
    \parbox{.48\textwidth}{\hspace{.8cm} \textbf{M stars}}
  }\\[-0.48cm]
  \mbox{
      \parbox{.48\textwidth}{
        \includegraphics[width=.48\textwidth,bbllx=60,bblly=10,bburx=620,bbury=440]{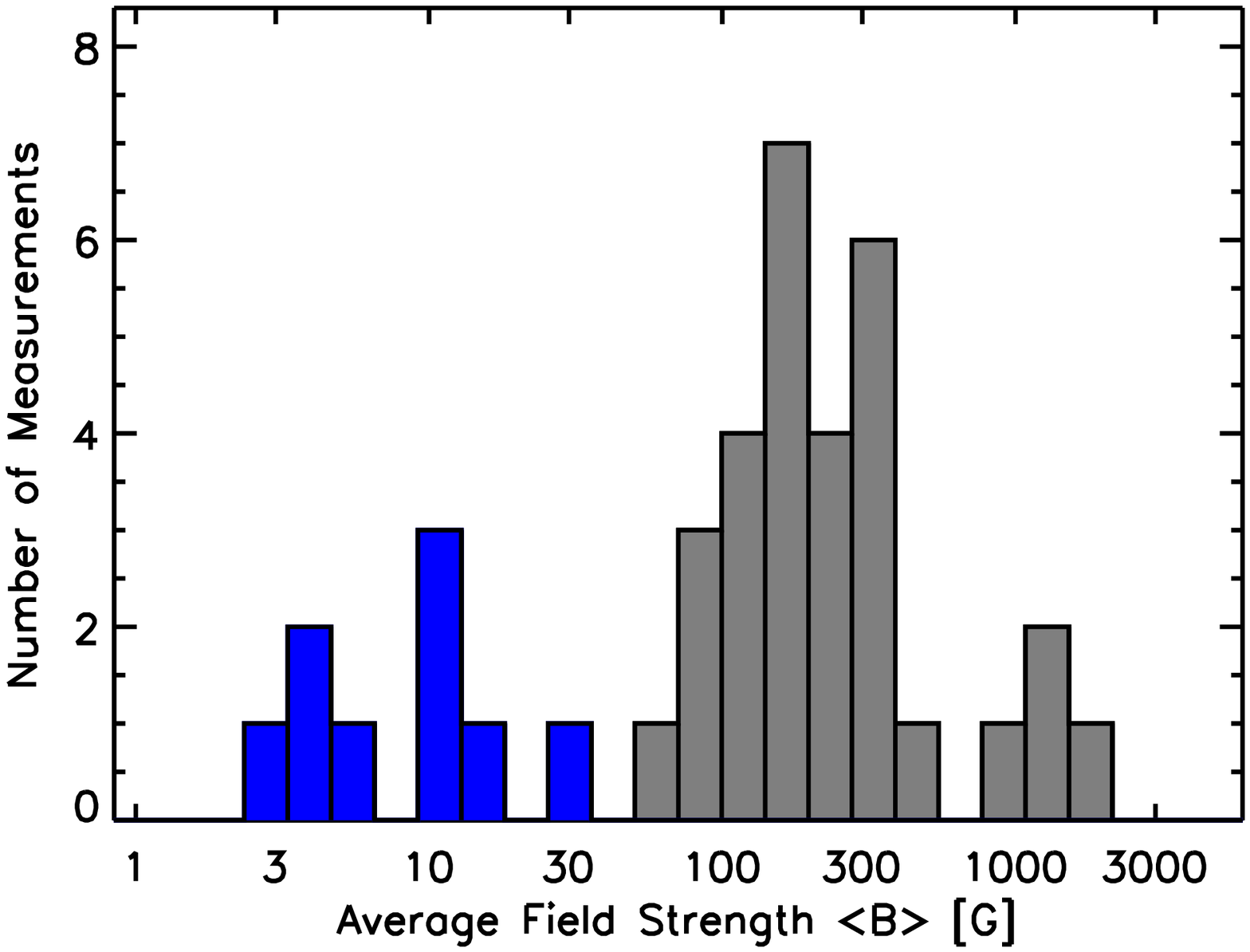}}
    \hspace{.04\textwidth}
    \parbox{.48\textwidth}{
      \includegraphics[width=.48\textwidth,bbllx=60,bblly=10,bburx=620,bbury=440]{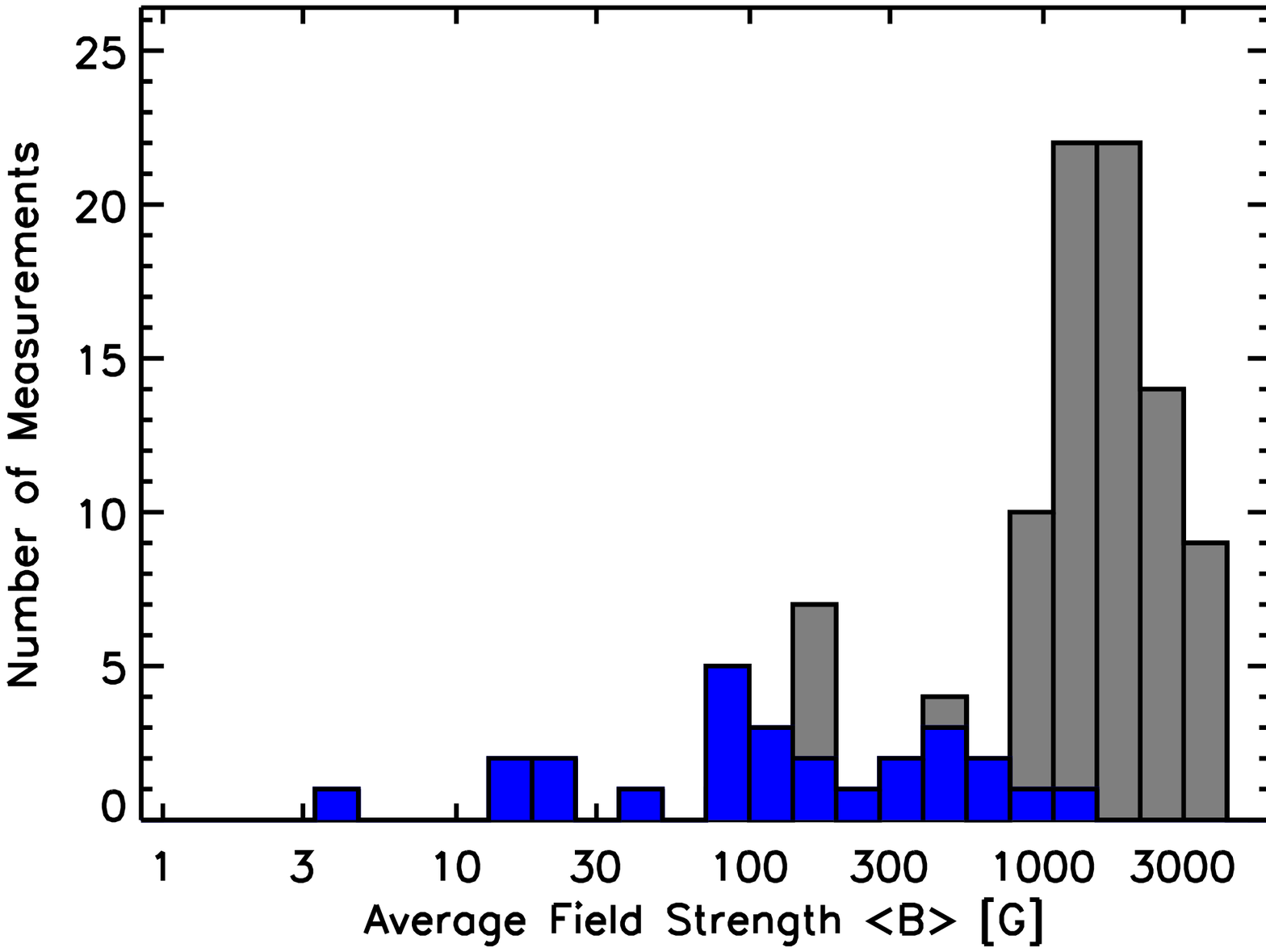}}
  }
  \caption{\label{fig:Distributions}Magnetic field measurements from
    Stokes~I (grey histograms) and Stokes~V (blue histograms) in G-
    and K-type stars (\emph{left} panel) and M stars (\emph{right}
    panel).}
\end{figure}

Most direct measurements of magnetic fields were carried out either in
Stokes~I or Stokes~V \cite[(but see Kochukhov et al., 2011)]{Kochukhov11}. The
observational systematics of the methods lead to significant biases that need
to be understood if we want to interpret the results. For example, Stokes~I
measurements have a hard time detecting weak magnetic fields in rapid
rotators, but they capture almost all field components. Stokes~V measurements,
on the other hand, can detect very small fields but cancellation of opposite
field directions can make significant field components invisible.

A collection of Stokes~I and Stokes~V average magnetic field
measurements is shown in Fig.\,\ref{fig:Distributions} \cite[(data
collection from Reiners, 2012)]{LivRev12}. Stokes~I measurements find
magnetic fields of several 100\,G and more, smaller fields cannot be
detected because of limited sensitivity. Stokes~V measurements in G-
and K-type stars are limited to field strengths of a few 10\,G, which
is probably because of cancellation effects. In M-stars, the detected
fields are significantly larger.

\begin{figure}
  \centering
      \includegraphics[width=\textwidth,bbllx=30,bblly=10,bburx=620,bbury=440]{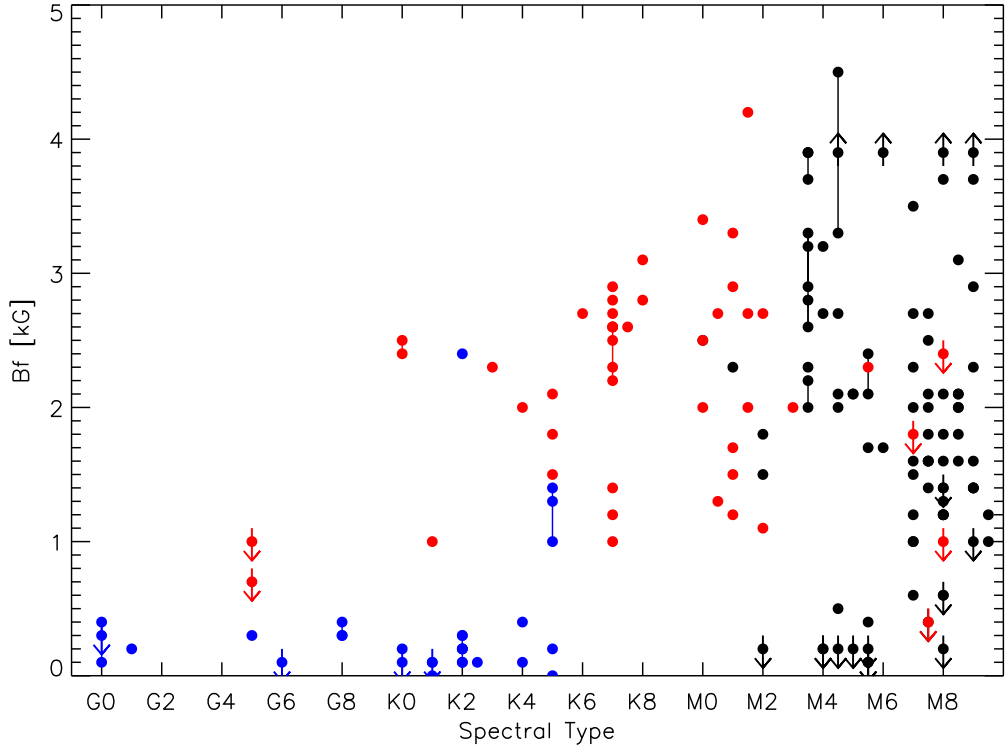}
      \caption{\label{fig:StokesI_all}Observations of Stokes~I average
        magnetic fields in different types of stars. \textit{Blue:} G- and
        K-dwarfs; \textit{black:} M-dwarfs; \textit{red:} pre-MS stars.}
\end{figure}

In Fig.\,\ref{fig:StokesI_all}, Stokes~I measurements for stars of spectral
types G0--M9 are shown as a function of spectral type. Field stars are shown
together with pre-main sequence stars. The magnetic fields found in late-type
stars are higher than in hotter stars, but not all late-type stars have strong
fields. The driver of magnetic field generation is rotation; the distribution
in Fig.\,\ref{fig:StokesI_all} reflects the distribution of rotation
velocities in stars of different spectral type and age. Furthermore, it
reflects the fact that strong fields are much more difficult to observe in G
stars than in M stars. The reason is the following: A field strength of ca.\
1000\,G can be expected in stars with Rossby numbers $Ro \approx
0.2$. According to Fig.\,\ref{fig:Rossby}, the equatorial rotation velocity of
a G2-star at this rotation rate is $v_{\mathrm{eq}} \approx
10$\,km\,s$^{-1}$. In an M0 star with the same Rossby number, the equatorial
rotation velocity is only $v_{\mathrm{eq}} \approx 3$\,km\,s$^{-1}$, and in an
M4 star we find $v_{\mathrm{eq}} \approx 1$\,km\,s$^{-1}$. The signature of
magnetic fields at low Rossby numbers is therefore much more obvious in
low-mass stars.

\end{document}